# Second ECOOP Workshop on Precise Behavioral Semantics (with an Emphasis on OO Business Specifications)

**Haim Kilov**[1], **Bernhard Rumpe**[2]

[1] Merrill Lynch, Operations, Services and Technology,
World Financial Center, South Tower
New York, NY 10080-6105, Haim_Kilov@ml.com
[2] Institut für Informatik, Technische Universität München
80333 Munich, Germany, Bernhard.Rumpe@in.tum.de

## 1   Motivation for the Workshop

Business specifications are essential to describe and understand businesses (and, in particular, business rules) independently of any computing systems used for their possible automation. They have to express this understanding in a clear, precise, and explicit way, in order to act as a common ground between business domain experts and software developers. They also provide the basis for reuse of concepts and constructs ("patterns") common to all – from finance to telecommunications –, or a large number of, businesses, and in doing so save intellectual effort, time and money. Moreover, these patterns substantially ease the elicitation and validation of business specifications during walkthroughs with business customers, and support separation of concerns using viewpoints.

Precise specifications of business semantics in business terms provide a common ground for subject matter experts, analysts and developers. All users of these specifications ought to be able to understand them. Therefore languages used to express such specifications should have precise semantics: as noted by Wittgenstein, "the silent adjustments to understand colloquial language are enormously complicated" [4]. (Not only English may be colloquial; graphical representations also may have this property[1].) If business specifications do not exist, or if they are incomplete, vague or inconsistent, then the developers will (have to) invent business rules. This often leads to systems that do something quite different from what they were supposed to do.

Business specifications are refined into business designs ("who does what when"), from where creation of various information system (software) specifications and implementations based on a choice of strategy and – precisely and explicitly specified! – environment, including technological architecture, are possible. In this con-text, precision should be introduced very early in the lifecycle, and not just in coding, as it often happens. In doing so, "*mathematics is not only useful for those who understand Latin, but also for many other Citizens, Merchants, Skippers, Chief mates, and all those who are interested*" ( Nicolaus Mulerius (1564-1630), one of the first three Professors of Groningen University).

---

[1] Probably, the most serious problem in this context is the usage of defaults and "meaningful" names. These are highly context-dependent and usually mean (subtly or not) different things for different people, including writers and readers. As a result, a possible warm and fuzzy feeling instead of a precise specification may lead to disastrous results.

Precise specification of semantics – as opposed to just signatures – is essential not only for business specifications, but also for business designs and system specifications. In particular, it is needed for appropriate handling of viewpoints which are essential when large and even moderately sized systems, both business and computer, are considered. Viewpoints exist both horizontally – within the same frame of reference, such as within a business specification – and vertically – within different frames of reference. In order to handle the complexity of a (new or existing) large system, it must be considered, on the one hand, as a composition of separate viewpoints, and on the other hand, as an integrated whole, probably at different abstraction levels. This is far from trivial.

Quite often, different names (and sometimes buzzwords) are used to denote the same concept or construct used for all kinds of behavioral specifications – from business to systems. "The same" here means "having the same semantics", and thus a good candidate for standardization and industry-wide usage. Various international standardization activities (such as the ISO Reference Model of Open Distributed Processing and OMG activities, specifically the more recent ones around the semantics of UML, business objects, and other OMG submissions, as well as the OMG semantics working group) are at different stages of addressing these issues. OMG is now interested in semantics for communities of business specifications, as well as in semantic requirements for good business and system specifications. Again, mathematics provides an excellent basis for unification of apparently different concepts (with category theory being a good example); and the same happens in science ("laws of nature"), linguistics, and business (for example, the Uniform Commercial Code in USA).

It is therefore the aim of the workshop to bring together theoreticians and practitioners to report about their experience with making semantics precise (perhaps even formal) and explicit in OO business specifications, business designs, software and system specifications. This is the 8th workshop on these issues; we already had 7 successful workshops, one at ECOOP and six at OOPSLA conferences. During workshop discussions, reuse of excellent traditional "20-year-old" programming and specification ideas (such as in [1,2]) would be warmly welcomed, as would be reuse of approaches which led to clarity, abstraction and precision of such century-old business specifications as [3]. Experience in the usage of various object-oriented modeling approaches for these purposes would be of special interest, as would be experience in explicit preservation of semantics (traceability) during the refinement of a business specification into business design, and then into a system specification.

The scope of the workshop included, but was not limited to:

- Appropriate levels and units of modularity
- Which elementary constructs are appropriate for business and system specifications? Simplicity, elegance and expressive power of such constructs and of specifications.
- Using patterns in business specifications
- Making Use Cases useful
- Discovering concepts out of examples (Generalization techniques)
- Providing examples from specifications
- What to show to and hide from the users
- How to make diagram notations more precise
- Equivalence of different graphical notations: "truth is invariant under change of notation" (Joseph Goguen)
- Semantics above IDL
- Rigorous mappings between frames of reference (e.g. business and system specifications)
- Role of ontology and epistemology in explicit articulation of business specifications
- Formalization of popular modeling approaches, including UML
- On complexity of describing semantics

The remainder of this paper is organized as follows. We overview the workshop's presentations (this overview is certainly biased; but its draft was provided to all and updated by some participants) and present the workshop's conclusions. Finally, the bulk of the paper consists those abstracts that the workshop's authors submitted (after the workshop) for inclusion. (Some authors did not submit any abstracts.)

The Proceedings of our workshop were published [6].

## 2 Overview

Starting almost without delay, the up to 24 participants have been viewing an interesting workshop. With even some visitors dropping in from the main conference, the workshop was an interesting event, at least as successful as its predecessors at previous years' OOPSLA and ECOOP conferences.

17 presentations gave an interesting overview of current and finished work in the area covered by this workshop. The workshop was organized in the following three sections:

**UML Formalization and Use**

*Kevin Lano* described translation of UML models to structured temporal theories, with the goal of enabling various developments that a UML user will typically do.

The user, however, does not need to use (or even see) the underlying framework. This led to a discussion about appropriate ways of showing the formalism to the users (e.g., by translating the formulas into rigorous English). *Roel Wieringa* emphasized the need to make explicit the methodological assumptions usually left implicit in UML formalization. (It was noted that UML formalizations tend to evolve into Visual C++.) The need to distinguish between the essential ("logical", determined by the external environment) and the implementation (determined by an implementation platform) was clearly emphasized. This essential decomposition results in circumscription of design freedom, making requirements explicit, being invariant under change of implementation, and providing explicit design rationale. *Claudia Pons* presented a single conceptual framework (based on dynamic logic) for the OO metamodel and OO model. UML—as a language – was considered as just a way to represent the axioms, so that the approach can be used for any two-level notation. *Tony Simons* presented a classification and description of 37 things that don't work in OO modeling with UML, and asked whether an uncritical adoption of UML is a "good thing". The presentation was based on experience with real development projects. Problems of inconsistency, ambiguity, incompleteness and especially cognitive misdirection (drawing diagrams, rather than modeling objects) were illustrated. UML diagrams mixed competing design forces (e.g. both data and client-server dependency, both analysis and design perspectives), which confused developers.

**Business and other Specifications**

*Offer Drori* showed how requirements for an information system were described using hypertext in industrial projects. He emphasized the need to bridge the gaps between the end user and the planner, and between the planner and programmer. He also stressed the need to understand (and make explicit) the "good attributes" of the existing system that are often ignored (remain implicit) in information management. *Bruce Siegel* compared a business specification approach for two projects – a rules-based and a Web-based one. Mixed granularity of requirements (very detailed datatype-like vs. very top-level understanding) was mentioned. The ANSI/IEEE SRS standard was used; use cases (which should have used pre- and postconditions) helped to define system boundaries; and loosely structured text was used to document screen functionality. State-based systems should specify pre- and postconditions for external system interactions. *Haim Kilov* described the business of specifying and developing an information system, from business specification, through business design and system specification, and to system implementation. The realization relationship between these stages was precisely defined, leading to clear traceability both "up" and "down", with an explicit emphasis on the (business, system, and technological) environment and strategy. Different realization variants and the need to make explicit choices between them were also noted. *Ira Sack* presented a comprehensive specification of agent and multi-agent knowledge including an epistemic ladder of abstraction. A hierarchy of knowledge types was developed using information modeling [14]; it clearly showed the power of precision. With information modeling, manage-ment people without specialized knowledge were able to understand epistemology in a business context. *Fatma Mili* described elicitation, representation and enforcement of automobile-related business rules. The need to find and gracefully update business rules (in collections of thousands of them) – without rewriting the whole system – was specifically noted. Reuse was made of invariants (what), protocols (how), and methods (when). Constraints are context-dependent, and a constraint should be attached to the smallest context possible. *Angelo Thalassinidis* described a library of business specifications (using information modeling [14]) for pharmaceutical industry; it was shown to business people who understood and appreciated the specifications. The goal was to enhance communication and to cope with change. It was proposed to use information modeling as a common approach to describe both the industry and the strategy at different layers (e.g., for industry – from the world through industries and corporations to divisions). "We have the HOW – information modeling; we need research as to WHAT to model." *Laurence Philips* described precise semantics for complex transactions that enable negotiation, delivery and settlements. He emphasized the need to be explicit about what is important and what can be ignored. Semantics was approximated by translation from "feeders" int o " interlingua" (that nobody externally uses) – this was the hardest step – added by "opportunistic" conflict resolution. Building code from "interlingua" i s easy assuming that infrastructure is in place. *Birol Berkem* described traceability from business processes to use cases. Since objects are not process-oriented, reusable "object collaboration units" are needed. A behavioral work unit consists of a dominant object and its contextual objects. Traceability between and within process steps and towards software development was noted.

**Formalization**

*Bernhard Rumpe* presented a note on semantics specifically mentioning the need to integrate different notations. Semantics is the meaning of a notation; "if you have a description of the syntax of C++ you still don't know what C++ does". Therefore "semantics is obtained by mapping a notation I don't know to a notation I do know" (and a notation is needed for the mapping itself). *Luca Pazzi* described statecharts for precise behavioral semantics and noted that behavior shapes critically the structure of the domain. Events compose to higher-level events. They denote state changes, are not directed, and should not be anthropomorphical. *Veronica Arganaraz* described simulation of behavior and object substitutability. Objects were described using Abadi-Cardelli imp$\varsigma$-calculus. A commuting diagram for a simulation relation was demonstrated. *Zoltan Horvath* presented a formal semantics for internal object concurrency. Proofs are done during, and not after, the design steps. To define a class, the environment of the system, and the user, are explicitly included, and so is the specification invariant.

Small discussions after each presentation and a larger discussion at the end allowed not only to clarify some points, but also to define points of interesting future research directions, and – most importantly – draw conclusions from this workshop.

## 3   Results of this Workshop

In order not to start from scratch, we started with the conclusions of our previous Workshop at ECOOP'97 [5]. And the participants came up with the following results. Most of them where accepted unanimously, some – only by majority (the latter are marked by *):

- Distinguish between specification and implementation
- Make requirements explicit
- Specifications are invariant under change of implementation
- Provide design rationale
- Distinguish between specification environment and implementation environment

- Specifications are used by specification readers and developers for different purposes
    - How can the users assert/deny the correctness of specifications – what representation to use?
    - Common ontology is essential:*
        - Ontology is the science of shared common understanding
        - Reality changes; and ontological frameworks need to change together with reality (Example – with the invention of surfboards, the legal question arose: is a surfboard a boat?)

- Cognitive problems in using (writing, reading) a notation may break a system*.

- A business specification may be partially realized by a computer system and partially by humans
- State-based systems should specify at least pre- and post-conditions for external system interactions; however, pre- and post-conditions may not be sufficient
- Different fragments ("projections", aspects) of system architecture should be visible to different kinds of (business) users

- A notation, to be usable, should be:
    - Understandable (no cognitive problems)
    - Unambiguous (* here no general agreement could be achieved. The authors believe that this mainly comes from a misunderstanding: "precise" is not the same as "detailed")
    - Simple (not "too much stuff")

- Different notations or representations are appropriate for different purposes (users)
- Reusable constructs exist everywhere:
  - in software specifications and implementations
  - in business specifications
  - in business design
  - Abstractions are also applicable throughout the whole development

- Composition exists not only for things but also for operations, events, etc. and often means a higher level of abstraction (Composition is not just 'UML-composition" but a concept in much wider use)

Points that have been discussed about, but no (common agreed) conclusions have been drawn:
- Common semantics, what it means and how to build? Comparison to mathematics, and other engineering disciplines.
- Is software engineering an engineering discipline?
- Are there some semantic basics and of what nature are they?

# Presentation abstracts

## 4 Traceability Management From 'Business Processes' to 'Use Cases' (*Birol Berkem*)

The goal of this work is to evaluate the applicability of the UML's activity diagram concepts for the business process modeling needs and to make a proposal for extensions to these concepts in order to define **formal traceability rules** from business processes to use cases. Robustness, testability, and executability of the business specifications appear as direct results of these extensions.

Object Oriented Business Process Modeling may be considered as the backbone tool for the Business Process Management, since it plays an important role for the design of the business process steps (business activities) around the right business objects and holds essential business rules for the system development. Managing a precise traceability from the business specifications layer to the system specifications layer is also useful to derive **process-oriented use cases**.

The usage of some basic elements of the UML's activity diagram doesn't allow this traceability between process steps. After presenting the characteristics of the UML's activity diagram in order to compare them with the Business Process Modeling needs, we remarked some lacks in the activity diagram concerning :
- the *management of the 'progress' for a given business process* :The **internal representation** of an activity (action state) is not provided to respond to 'why and how the action is performed' in order to model the goal and the emerging context ( i.e., what is the behavioral change that happens in the owning object of an action state to perform the action and what are **the responsibilities requested from other objects** participating in the action, allowing that way the measurement of the performances for a business process via its different stages).

- *the management of the object flows between activities* : There is no information about the **destination of the outputs** produced by an activity **inside** its corresponding action state. Indeed, any organizational unit that executes a process step should have the knowledge of the destination of its outputs depending on the status of completion of the activity. This requires the definition of relevant output objects **inside** each activity (action state), each one expressing the required responsibility.

In order to formalize objects responsibilities within business process, we proposed to make a zoom on each action state to discover participating objects then determine all the necessary information that output objects should carry out via their links to the target process steps. This should also *reduce the information fetch time* for the actors that use these target process steps.

Considering that business processes **must be directed by goals**, we have been conducted to apply to the UML's Activity Diagram the concept of 'Contextual Objects' where **objects are driven by goals**. This allows:

- a robustness in the implementation of **executable specifications** via the chronological forms (to execute, executing, executed) of the object behaviors then a precise response for the progression of a process,
- a formal way in the definition of the **behavioral state transition diagram** (an extension to the UML's state transition diagram) which represents the internal behavior of the 'action states' and their transition,
- an implicit definition of the **right business objects** that can be captured along the process modeling,
- finally, a formal way **to find out process-oriented 'use cases' and their 'uses / includes' relationships** using work units **or** actions states internals of a process step.

## 5 Definition of Requirements for an OODPM-Based Information System Using Hypertext (*Offer Drori*)

Information systems are developed along a time axis known as the system life cycle. This cycle comprises several stages, of which the principal ones are: initiation, analysis of the existing situation, applicability study, definition of the new system, design, development, assimilation, and maintenance. The system definition stage is effectively the stage in which the systems analyst summarizes the user's needs, and constitutes the basis for system design and development. Since this is a key stage, every effort must be made to ensure that all relevant issues are actually included, and that the requirements definition extracts the full range of user needs for the planned information system. The present article aims to describe a method for defining the requirements of an OODPM-based information system using hypertext, based on a HyperCASE computerized tool.

All elements of an information system can be controlled and supervised by its specification. The specification of the various elements is essential, and creating the link between the specification of the elements and the actual elements is vital. An integrative approach embracing all information system elements will enable confronting a large portion of the problems associated with information systems development.

OODPM - Object Oriented Design using Prototype Methodology is a system planning and design method that integrates the two approaches contained in its title.

OOPDM focuses primarily on system planning, but also addresses the business specification stage. According to this approach, user needs to be implemented in the future system must be studied, but time must also be dedicated to studying the current situation in order to complete the requirements definition. Experience has shown that users tends to focus on those needs that have not been met by the current system, and tend to ignore the parts of the system that have met their needs. Without a stage to examine the current situation, only partial definition of the requirements is likely to achieved.

In sum, with OODPM, the system analysis and planning process begins with a study of the current situation, but with a view to identifying the needs, rather than the study for its own sake. This means that a defined period of system planning time, proportional to the overall process, is assigned to the study of the current situation. This process ends with the business specifications, or as it is usually called, the business specifications for the new system.

System planning with OODPM is done by defining the activities required for the system. A system activity is defined as a collection of data and processes that deal with a defined subject and are closely linked. Data affiliated with a particular process should be collated in a natural manner; however, should there be tens of data pertaining to a specific process, secondary definition of sub-processes is desirable. One can also add to the activity definition the requirement that the user, on a single screen, process the data in a single, continuous action in a single sequence. A collection of user activities in a particular area with the relationships between them defines an information system. (Some of these user activities may be manual.)

The software crisis, which resulted in a crisis of confidence between information systems developers and users, can be resolved. There are many ways to go about this. This article focused on two such ways; one, the adoption of OODPM - a systematic, structured methodology for planning and developing information systems. Two, the use of a hypertext-based tool to create a superior requirements specification taking into account the future system users. The experience gained with this tool in both academic and 'real life' system s development environments points to positive results for this approach. Recently we also see CASE tools that are partially have hypertext-based tools for systems managing but they still need more comprehensive and profound attitude.

## 6 A Formal Semantics of Internal Object Concurrency (*Ákos Fóthi, Zoltán Horváth, Tamás Kozsik, Judit Nyéky-Gaizler, Tibor Venczel*)

We use the basic concepts of a relational model of parallel programming. Based on the concepts of a problem, an abstract program and a solution we give a precise semantics of concurrent execution of abstract data type operations. Our approach is functional, problems are given an own semantical meaning. We use the behavior relation of a parallel program which is easy to compare to the relation which is the interpretation of a problem.

In this paper we show a simple method to solve the complex problem of correct - i.e. surely adequate both to the specification and to the chosen representation - parallel implementation of abstract data types.

Finally the practical advantage of its use is shown on a concrete example.
Our model is an extension of a powerful and well-developed relational model of programming which formalizes the notion of state space, problem, sequential program, solution, weakest precondition, specification, programming theorem, type, program transformation etc. We formalize the main concepts of UNITY in an alternative way. We use a relatively simple mathematical machinery.

Here we do not consider open specifications. The specification is given for a joint system of the program and the environment. The programs are running on different subspaces of the state space of the whole system, forming a common closed system based on principles similar to the generalized action-oriented object model.

A generalization of the relational (e.g. not algebraic) model of class specification and implementation for the case of parallel programs is shown. Instead of using auxiliary variables for specifying objects in parallel environments we generalize the concept of type specification and type implementation. A difference is made between the specification and implementation of a data type as like as between problem and program. So we define the class specification consisting of the type value set, specification invariant, the specification of the operations and specification properties for the environment of the data type. Next we introduce the implementation of a class, which is build up of the representation, type invariant and the implementation of the operations. We give a relational semantics to an implementation being adequate to a class specification.

The semantics is used for a precise refinement calculus in problem refinement process and for verification of the correctness of the abstract parallel program. The verification of the correctness of the refinement steps and the correctness of the program in respect to the last specification may be performed by an appropriate temporal logic based verification tool in the future. The introduced model makes it possible to define operations that can run in parallel i.e. internal concurrency of objects is allowed. Nonterminating operations are allowed as well. The class set'n' cost is used to demo strate the applicability of this methodology.

## 7 The graph-based Logic of Visual Modeling and Taming Heterogeneity of Semantic Models (*Zinovy Diskin*)

The goal of the paper is to explicate some formal logic underlying various notational systems used in visual modeling (VM). It is shown that this logic is a logic of predicates and operations over arrow diagrams, that is, a special graph-based logic of *sketches*: the latter are directed multi-graphs in which some diagrams are marked with labels taken from a predefined signature. The idea and the term are borrowed from categorical logic, a branch of mathematical category theory, where sketches are used for specifying mathematical structures. Thus, VM-diagrams are treated as visual presentations of underlying (formal graph-based) sketch specifications. In this way the diversity of VM notations can be presented as a diversity of visualizations over *the same* specificational logic. This gives rise to a consistent and mathematically justified unification of the extremely heterogeneous VM-world.

The approach can be realized only within some formal semantic framework for VM. And indeed, in the paper it is outlined how basic constructs of conceptual modeling (like IsA, IsPartOf, various aggregation and qualification relationships) can be formally explicated in the framework of variable set semantics for sketches (see [7] for details).

In a wider context, the goal of the paper is to manifest the arrow style of thinking as valuable both in the practice of conceptual modeling and design and in stating their logically consistent foundations as well. VM-diagrams are to be thought of as high-level semantic specifications rather than graphical interfaces to relational and the like low-level schemas. The machinery of diagram predicates and operations proposed in the paper is intended to support this thesis on the technical logical level.

## 8 An Information Management Project: What to do when your Business Specification is ready (*Haim Kilov, Allan Ash*)

We present a business specification of the business of *developing an information system*.

Since different activities during an information management project emphasize different concerns, it makes sense to separate these concerns and, in particular, to use different names to denote them – business specification, business design, system specification, and system implementation. Each activity builds a description of the system from its viewpoint, which then is "*realized*" or moved to a more concrete (more implementation-bound) viewpoint, – by the next activity. The frames of reference of these activities have a lot of common semantics which is essential for being able to bridge the gap from businesses to systems.

The *Realization* relationship relates the "source" activity and the "target" activity which realizes it. A target is not uniquely determined by its source. There may be more than one business design that will realize a given business specification, for example. Generally, any number of realization variants may be generated which will result in multiple versions of the target activity, although only the "best" one will be chosen to be realized by the next activity.

A more detailed specification of this relationship is provided in the figure below.

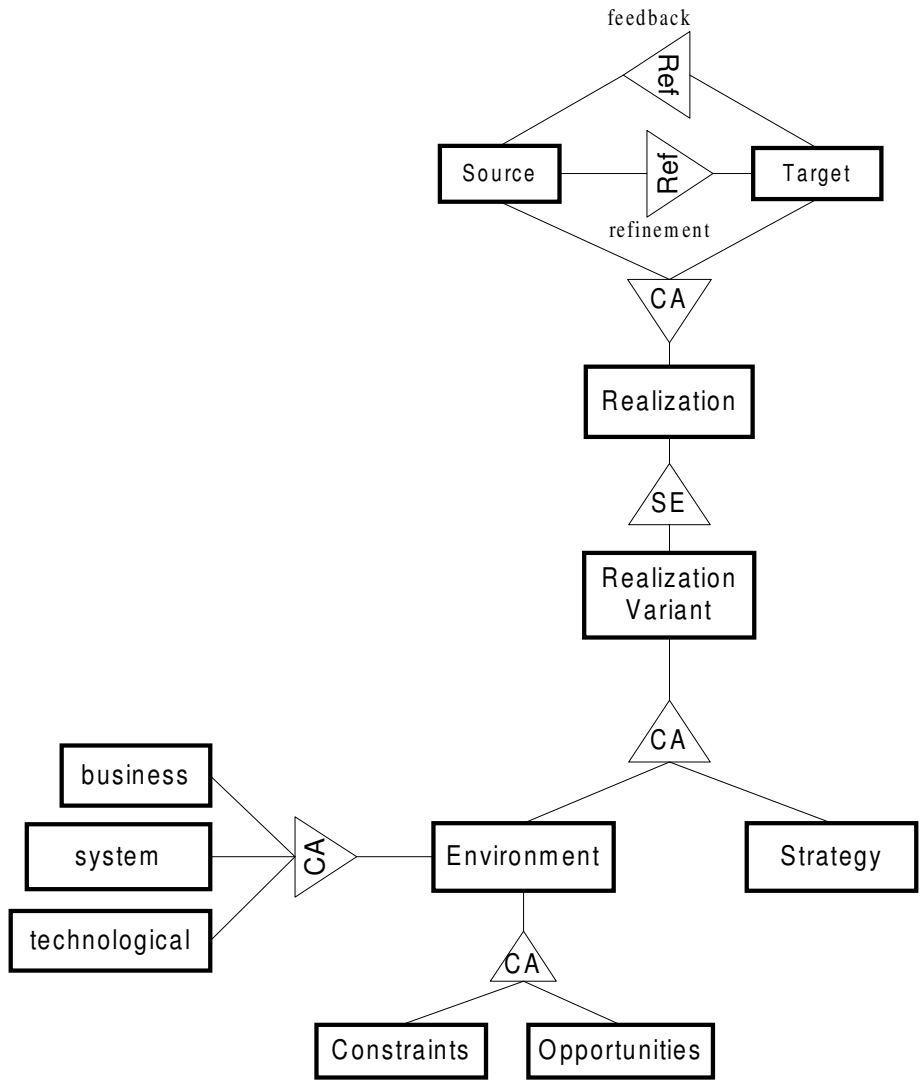

## 9 Formalising the UML in Structured Temporal Theories (*Kevin Lano, Jean Bicarregui*)

We have developed a possible semantics for a large part of the Unified Modelling Notation (UML), using structured theories in a simple temporal logic. This semantic representation is suitable for modular reasoning about UML models. We show how it can be used to clarify certain ambiguous cases of UML semantics, and how to justify enhancement or refinement transformations on UML models.

The semantic model of UML used here is based on the set-theoretic Z-based model of Syntropy [9]. A mathematical semantic representation of UML models can be given in terms of *theories* in a suitable logic, as in the semantics presented for Syntropy in [8] and VDM$^{++}$ in [11]. In order to reason about real-time specifications the more general version, Real-time Action Logic (RAL) [11] is used.

A typical transformation which can be justified using our semantics is source splitting of statechart transitions. We intend that such transformations would be incorporated into a CASE tool, so that they could be used in practical development without the need for a developer to understand or use the formal semantics.

## 10 Alchemy Is No Substitute For Engineering In Requirement Specification (*Geoff Mullery*)

Attempts to be "formal" in specifying computer systems are repeatedly reported as having failed and/or been too expensive or time-consuming. This has led some people to assert that the use of science in this context has failed and that we should instead concentrate on use of what are perceived as non-science based methods (for example approaches driven by sociological investigation).

In reality what has been applied from science is a limited subset based on mathematics (characterised by Formal Methods) or on pseudo mathematical principles (characterised by Structured Methods). There is more to science than this subset and failure to apply all aspects of science is arguably a prime reason for failure in previous approaches.

Science in general can be characterised in terms of Models and Disciplines. Models lead to the ability to characterise a proposed system in terms of deductions and theorems based on the model definition or its application to description of a specific problem. Disciplines produce, make use of and evaluate models in varying ways, depending on the primary thrust of their sphere of interest.

Models may be Abstract (not representing on the observable universe, though possibly derived from it) or Real World (based on the observable universe and representing a subset of it). A Well Formed model has an internally consistent definition according to mathematically accepted criteria-derived deductions and theorems are likely to be correspondingly consistent. A model which is not well formed is likely to demonstrate inconsistencies in deductions and theorems derived from its use.

In computing specifications are models and represent only a subset of a proposed system and its environment so it must be incomplete (otherwise it would be a clone of the proposed system + environment). Also, even for well formed models, it is impossible to be 100% certain that all deductions/theorems are correct, so a model is suspect even over its domain of applicability.

The Pure Science discipline defines models, but rarely worries about their application to the Real World. The Applied Science discipline defines models based on the Real World or maps part of the Real World onto a model defined by Pure Science. It is frequently the case that Applied Scientists are more interested in producing the models than they are in ensuring that they are well formed.

The Engineering discipline tests and uses models in the Real World, discovering areas of applicability and margins for safety of application. The Control Engineering discipline facilitates model application, evaluation and improvement by looking for divergence between model predictions (via deductions and theorems) and behaviour observed when used in a Real World mapping.

It is in Engineering and Control Engineering that the computer industry has made little practical use of science. Evaluation of methods and their underlying models has been based on anecdotal evidence, marketing skills, commercial pressures and inappropriate, frequently biased experiments. Method and model advocacy has been more akin to alchemy than to science. The alternative of ignoring science and using methods with a series of ad hoc pseudo models is merely one which accentuates alchemy - with the result that we are in danger of getting only more of the same.

Nevertheless it is also clear that, since all models represent only a subset of the universe of interest, there is no guaranteed profit in assuming that all that is needed are well formed models. What is needed is an attempt to integrate the use of well formed and ad hoc models, with co-operation in the process of translating between the models - and in both directions, not just from ad hoc to formal.

And finally, in setting up such a co-operative venture the critical thing to ensure that the venture really works and can be made gradually to work even better is to apply the disciplines of Engineering and the more specialised Control Engineering as characterised here. That is the minimum requirement for the achievement of improved systems development and that is the basis of the Scientific Method.

## 11 Part-Whole Statecharts for Precise Behavioral Semantics (*Luca Pazzi*)

The work presented by Luca Pazzi suggested that a close and critical relationship exists between the behavioral and structural knowledge of complex engineering domains. It may be observed that most of the formalisms for representing aggregate entities present a tendency towards either an implicit or explicit way of representing structural information. By the *implicit approach*, a complex entity is modeled through a *web* of references by which the component entities refer to one another.
This is typical, for example, of the object-oriented approach, which models an associative relationship between two objects, for example car A towing trailer B, by an object reference from A to B. This way poorly reusable abstractions results (for example car A becomes, structurally, a tower). The counterpart is represented by the *explicit approach*, where the emphasis is on the explicit identification of a whole entity in the design, be it an aggregate or a regular entity. The claim is that such identification may be driven by analysing the associative knowledge, i.e. usually behavioral relationships, observed in the domain. *Behavior* contributes thus in determining *additional structure* in the domain and such identification impacts *critically* on the overall quality of the modeling. Behavioral specifications play thus a mayor role in committing a modelling formalism towards the explicit approach.

## 12 Integrating Object-Oriented Model with Object-Oriented Metamodel into a single Formalism (*Claudia Pons, Gabriel Baum, Miguel Felder*)

Object oriented software development must be based on theoretical foundations including a conceptual model for the information acquired during analysis and design activities. The more formal the conceptual model is, the more precise and unambiguous engineers can be in their description of analysis and design information.

We have defined an object-oriented conceptual model [13] representing the information acquired during object-oriented analysis and design. This conceptual model uses explicit representation of data and metadata into a single framework based on Dynamic Logic, allowing software engineers to describe interconnections between the two different levels of data.

We address the problem of gaining acceptance for the use of an unfamiliar formalism by giving an automatic transformation method, which defines a set of rules to systematically create a single integrated dynamic logic model from the several separate elements that constitute a description of an object-oriented system expressed in Unified Modeling Language.

The intended semantics for this conceptual model is a set of states with a set of transition relations on states. The domain for states is an algebra whose elements are both data and metadata. The set of transition relation is partitioned into two disjoint sets: a set of transition representing modifications on the specification of the system (i.e. evolution of metadata), and a set of transition representing modifications on the system at run time (i.e. evolution of data).

The principal benefits of the proposed formalization can be summarized as follows:
- The different views on a system are integrated into a single formal model. This allows one to define rules of compatibility between the separate views, on syntactical and semantical level.
- Formal refinement steps can be defined on model.
- This approach introduces precision of specification into a software development practice while still ensuring acceptance and usability by current developers.
- The model is suitable for describing system evolution; it is possible to specify how a modification made to the model impacts on the modeled system. By animating the transition system defined by the formal specification it is possible to simulate the behavior of the specified system and also it is possible to analyze the behavior of the system after evolution of its specification (either structural evolution or behavioral evolution or both).
- The model is suitable for formal description of reuse contracts, reuse operators, design patterns and quality assessment mechanisms.

## 13 Simulation of Behaviour and Object Substitutability (*Maria José Presso, Natalia Romero, Verónica Argañaraz, Gabriel Baum, and Máximo Prieto*)

Many times during the software development cycle, it is important to determine if an object can be replaced by another. In an exploratory phase, for example, when prototyping a system or module simple objects with the minimum required behaviour are defined. In a later stage they are replaced by more refined objects that complete the functionality with additional behaviour. The new objects must emulate the portion of functionality already implemented, while providing the implementation for the rest of it.

During the evolution of a system, to improve the performance of a module for example, there is also the need to change some objects for others with a more efficient implementation. In this case we need to assess that the replacing object has exactly the same behaviour as the replaced, retaining the functionality of the whole system unchanged.

What we need is a substitutability relation on objects that can be used to determine if an object can be replaced by another. We may say that an object can substitute another if it exhibits "at least the same behaviour". Or we could strengthen the condition, and ask it to exhibit "exactly the same behaviour". In both interpretations of substitutability, the behaviour of the first object must be emulated by the new object, but in the first one the new object is allowed to have extra functionality.

The purpose of this work is to define the notion of two objects having the same behaviour. We have discussed two different relations that characterise the idea. In order to have a useful and clear notion of these relations we must define them rigorously. Such a rigorous formulation allows us to precisely state when two objects have the same behaviour, so that one can replace the other while ensuring the preservation of the semantics of the whole system.

We take the simulation and bisimulation techniques, widely used in semantics of concurrent systems, as formal characterisations of the relation of having the same behaviour.

As a formal framework to represent objects we use the imp$\varsigma$-calculus of Abadi and Cardelli, which is a simple, object based calculus, with an imperative semantics. The imperative semantics allows to model appropriately some key issues of object oriented programming languages such as state, side effects and identity of objects.

The simulation and bisimulation for objects in the calculus are defined using a labelled transition system based on the messages that objects understand, and takes into account the possible side effects of message passing, present in the imperative calculus semantics.

We propose the defined simulation relation can be used to formally characterise the idea of one object having "at least the same behaviour" as another. Similarly, bisimulation is defined to capture the idea of an object having "exactly the same behaviour" as another.

## 14 A Note on Semantics with an Emphasis on UML (*Bernhard Rumpe*)

> "In software engineering people often believe a state is a node in a graph and don't even care about what a state means in reality."
>
> David Parnas, 1998

This note clarifies the concept of syntax and semantics and their relationships. Today, a lot of confusion arises from the fact that the word "semantics" is used in different meanings. We discuss a general approach at defining semantics that is feasible for both textual and diagrammatic notations and discuss this approach using an example formalization. The formalization of hierarchical Mealy automata and their semantics definition using input/output behaviors allows us to define a specification, as well as an implementation semantics. Finally, a classification of different approaches that fit in this framework is given. This classification may also serve as guideline when defining a semantics for a new language.

## 15 Towards a Comprehensive Specification of Agent and Multi-agent Knowedge Types in a Globalized Business Environment (*Ira Sack, Angelo E. Thalassinidis*)

We produce a detailed level specification based on information modeling as defined in [14] refined by a frame-based semantics presented in [15] to precisely specify various types of agent and multi-agent *knowledge types* in a globalized business environment. Our approach results in a definition of agent and multi-agent knowledge types based on the collective works of the authors cited in the references. The highest level of specification consists of information molecules (displayed as Kilov diagrams) that show *concept linkage* between epistemic notions such as agent knowledge types and possible worlds. It is our belief that information modeling is preferable to object-oriented modeling when specifying very high level abstractions (super-abstractions?) such as possible worlds, knowledge types, and information partitions and the linkages (known in information modeling as *associations*) between them. It is also a reasonable and appropriate means to present and "socialize" notions which are increasingly becoming focal points for new approaches to information and business system design (e.g., market oriented systems, intelligent agents, negotiation systems).

Whereas Wand and Wang have addressed the issue of data quality of an information system from an ontological perspective premised on a one-to-one correspondence between a set of information states and a set of states representing real-world perceptions, they did not specifically address the issue of uncertainty. Using information modeling and Aumann structures we have extended the work presented in [16] to include and model uncertainty in agent perception within a multiple agent perspective.

We have also created an information molecule that includes five multi-agent knowledge subtypes: distributed knowledge, some agent knows, mutual knowledge, every agent knows that every agent knows, and common knowledge. These five subtypes are assembled to form an *Epistemic Ladder of Abstraction (ELA)* – an information molecule subject to a set of logical constraints that order the various knowledge types in terms of their levels of abstraction – with common knowledge sitting at the top.

## 16 ` 37 Things that Don' t Work in Object-Oriented Modelling with UML (*Anthony J H Simons, Ian Graham*)

The authors offer a catalogue of problems experienced by developers using various object modelling techniques brought into prominence by the current widespread adoption of UML standard notations. The problems encountered have different causes, including: ambiguous semantics in the modelling notations, cognitive misdirection during the development process, inadequate capture of salient system properties, features missing in supporting CASE tools and developer inexperience. Some of the problems can be addressed by increased guidance on the consistent interpretation of diagrams. Others require a revision of UML and its supporting tools. The 37 reported problems were classified as: 6 inconsistencies (parts of UML models that are in self-contradiction), 9 ambiguities (UML models that are underspecified, allowing developers to interpret them in multiple ways), 10 inadequacies (concepts which UML cannot express adequately) and 12 misdirections (cases where designs were badly conceptualised, or drawn out in the wrong directions). This last figure is significant and alarming. It is not simply that the UML notation has semantic faults (which can be fixed in later versions), but rather that the increased prominence given to particular analysis models in UML has in turn placed a premium on carrying out certain kinds of intellectual activity, which eventually prove unproductive. Our analysts enthusiastically embraced the new use-case and sequence diagram approaches to conceptualising systems, generating control structures which our designers could not (and refused to) implement, since they did not map onto anything that a conventional software engineer would recognise.

Similar disagreements arose over the design interpretation of analysis class diagrams due to the tensions between the data dependency and client-supplier views; and the place and meaning of state and activity diagrams. Most problems can be traced back to the awkward transition between analysis and design, where UML' s universal philosophy (the same notation for everything) comes unstuck. Modelling techniques that were appropriate for informal elicitation are being used to document hard designs; the same UML models are subject to different interpretations in analysis and design; developers are encouraged to follow analytical procedures which do not translate straightforwardly into clean designs. UML is itself neutral with respect to good or bad designs; but the consequences of allowing UML to drive the development process are inadequate object conceptualisation, poor control structures and poorly-coupled system designs.

## 17 Association Semantics in Medical Terminology Services (*Harold Solbrig*)

This paper describes some of the interesting issues encountered during the development of a standard interface specification for accessing the content of medical terminologies. It begins by briefly describing some of the common medical coding schemes and outlines some of the motivations for developing a common interface to access their content. It then proceeds to describe one such specification, the Lexicon Query Services (LQS) interface specification, which was recently adopted by the Object Management Group.

Medical terminology systems frequently define the concepts behind the terminology in terms of their associations with each other. Concepts are frequently defined as a *type of* another concept, or *broader than* or *narrower than* another concept in scope. Various forms of subtype and subclass associations occur as well as associations like *contains, is composed of,* etc. As the meaning behind a given medical term is dependent on the communication of these associations, it was determined that a formalization of the association semantics would be necessary if this specification was to be generally useful.

This paper describes some of the issues that were encountered when the authors attempted to apply a form of association semantics used in object-oriented modeling to the semantics of medical terminology associations.

## 18 Building the Industry Library – Pharmaceutical (*Angelo E. Thalassinidis, Ira Sack*)

Both OO and business research communities have not yet operationalized nor even formalized high-level business concepts such as strategy, competition, market forces, product value, regulations, and other "soft" business notions. This may be attributed to two main reasons: a) OO researchers are using an incremental approach in building libraries that is fundamentally bottom-up — believing it is only a matter of time until they can address high-level business concepts; and b) Researchers from the business strategy side have never attempted to formalize these concepts due to the difficulties engendered by their differing backgrounds (psychology, economics, etc.) or the different audiences they must address (CEOs will not read a detail library).

This paper constitutes the first of a series that present an ongoing effort in building "Business Industry Libraries" (BIL, for short) using modeling constructs introduced in [14]. A BIL will model the specific characteristics of an industry. The BIL will be accompanied by a "Business Strategy Library" that the authors have started working on in [18,19,20], an "Organizational Theory Library" whose foundation is presented in [17], and other constructs. BIL will assist in illuminating difficult business considerations facilitated by the employment of as much precision as the vagaries, instabilities, etc., allow.

The type of information that the business library should be maintaining is still being researched and will be presented in an upcoming paper.

This paper accomplishes the analysis of the pharmaceutical industry; one of the most complicated industries. The pharmaceutical industry is generally viewed by analysts as the composition of *Human-use*, *Animal-use*, *Cosmetic-use*, and *Food-use* products. The Human-use products are drugs developed to address human health needs; the Animal-use products are drugs developed to address either animal health needs or human needs from animal products; Cosmetic-use products are drugs developed to address human esthetic needs; and; Food-use products are drugs developed to address human, animal, or even plant needs. The pharmaceutical industry is also international by nature. In many countries the pharmaceutical industry has a relationship with its customers that is unique in manufacturing. The industry provides drugs, the physician decides when to use them, the patient is the consumer, and the bill is predominantly paid by a private or national insurance subject to a co-payment.

## 19 Formalizing the UML in a Systems Engineering Approach (*Roel Wieringa*)

This discussion note argues for embedding any formalization of semiformal notations in a methodology. I present a methodological framework for software specification based on systems engineering and show how the UML fits into this framework. The framework distinguishes the dimensions of time (development strategy and history), logic (justification of the result), aspect of the delivered system, and aggregation level of the system. Aspect is further decomposed into functions, behavior and communication. Development methods offer techniques to specify these three aspects of a software system. The UML offers use case diagrams to specify external functions, class diagrams to specify a decomposition, statecharts to specify behavior and sequence and collaboration diagrams to specify communication.

Next, an essential modeling approach to formalizing the UML within this framework is argued. This means that the we should define an implementation-independent decomposition, that remains invariant under changes of implementation. Finally, a transition system semantics for the UML is discussed, that fits within the semantic modeling approach. The semantics models an object system as going through a sequence of steps, where each step is a finite set of actions performed by different objects. Objects communicate by means of signals. This semantics has been formalized as a transition system semantics.

No formal details are given, but references are given to places where these can be found.

## References

1. E.W.Dijkstra. On the teaching of programming, i.e. on the teaching of thinking. In*: Language hierarchies and interfaces (Lecture Notes in Computer Science, Vol. 46),* Springer Verlag, 1976, pp. 1-10.